\UseRawInputEncoding
\documentclass[a4paper]{jpconf}
\usepackage{graphicx}
\RequirePackage{inputenc,crop,graphicx,amsmath,array,color,amssymb,flushend,stfloats,amsthm,chngpage,times,datetime,parskip,epstopdf}
\begin{document}
\title{Search for dark matter with IACTs and the Cherenkov Telescope Array}

\author{Aldo Morselli $^1$, on behalf of the CTA Consortium}
\address{$^1$ INFN Roma Tor Vergata, Via della Ricerca Scientifica 1   00133 ROMA, ITALY }

\ead{aldo.morselli@roma2.infn.it}

\begin{abstract}
In the last decades an incredible amount of evidence for the existence of dark matter (DM) has been accumulating. At the same time, many efforts have been undertaken to try to identify what dark matter is made of.
 Indirect searches look at places in the Universe where dark matter is known to be abundant and seek for possible annihilation or decay signatures. Indirect searches with the Fermi Gamma ray Space Telescope and Imaging Atmospheric Cherenkov Telescopes (IACTs) are playing a crucial role in constraining the nature of the DM particle through the study of their annihilation into gamma rays from different astrophysical structures. In this talk I will review 
the status of the search with IACTs and I will describe the sensitivity projections for dark matter searches on the various targets taking into account the latest instrument response functions expected for the Cherenkov Telescope Array (CTA) together with estimations for the systematic uncertainties from diffuse astrophysical and cosmic-ray backgrounds.
\end{abstract}

\section{Introduction}
High-energy phenomena in the cosmos, and in particular processes leading to the emission of gamma rays in the energy range 10 MeV - 100 TeV, play a very special role in the understanding of our Universe. This energy range is indeed associated with non-thermal phenomena and challenging particle acceleration processes. The Universe can be thought as a context where fundamental physics, relativistic processes, strong gravity regimes, and plasma instabilities can be explored in a way that is not possible to reproduce in our laboratories. High-energy astrophysics and atmospheric plasma physics are indeed not esoteric subjects, but are strongly linked with our daily life. Understanding cosmic high-energy processes has a large impact on our theories and laboratories applications. The technology involved in detecting gamma rays is challenging and drives our ability to develop improved instruments for a large variety of applications.
In the last decades a vast amount of evidence for the existence of dark matter has been accumulated. At the same time, many efforts have been undertaken to try to identify what dark matter is made of. Indirect searches look at places in the Universe where dark matter is believed to be abundant and seek for possible annihilation or decay signatures. 
At high energies, the Cherenkov Telescope Array (CTA) represents the next generation of imaging Cherenkov telescopes and, with one array in the southern hemisphere and one in the northern hemisphere, will be able to observe all the sky with unprecedented sensitivity and angular resolution from 20 GeV up to 300 TeV. The CTA Consortium will undertake an ambitious program of indirect dark matter searches for which we report here the prospects.

\begin{figure}
  \centering
 \includegraphics[width=0.80\textwidth]{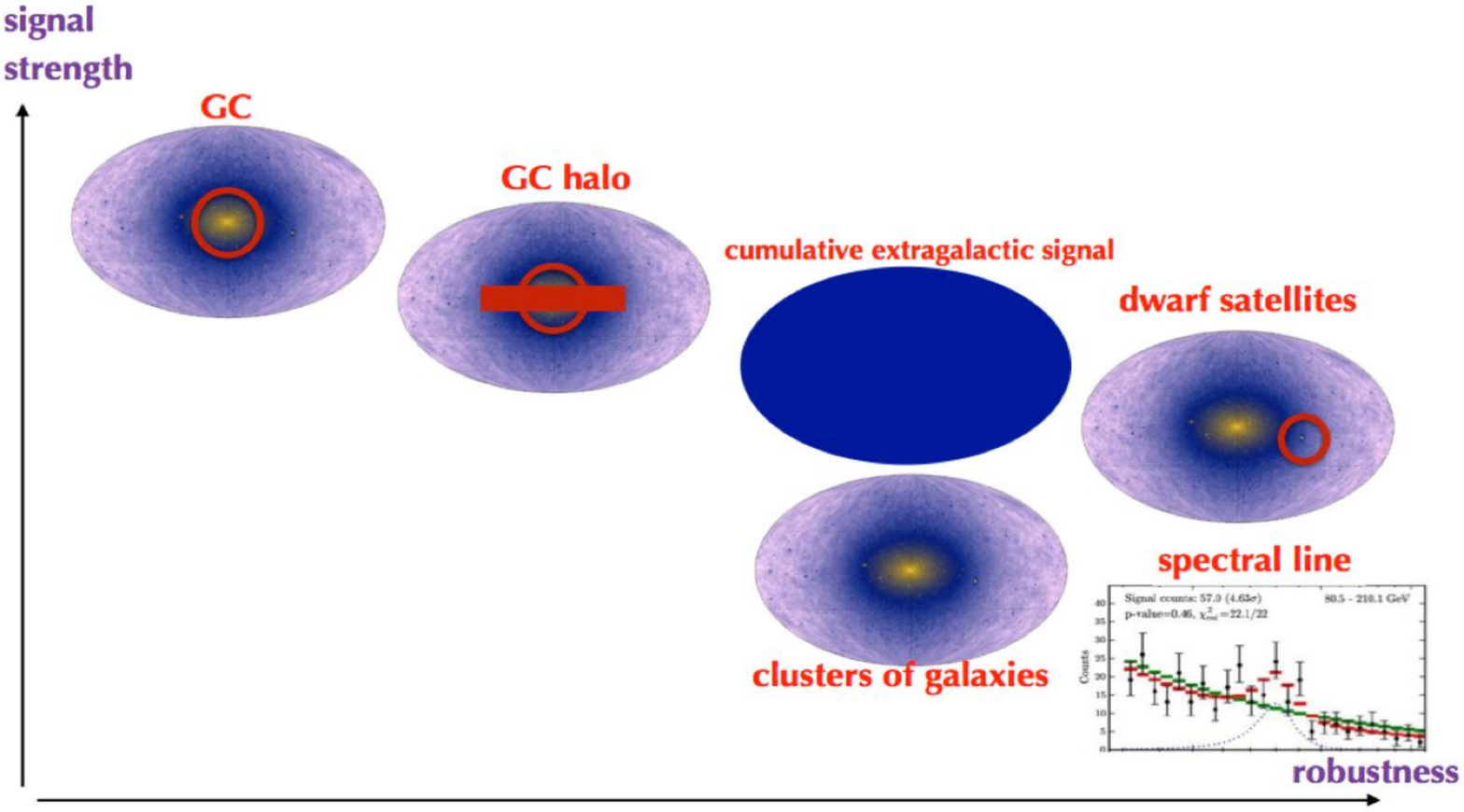}
  \caption{ Signal strength versus robustness of the various possible dark matter signals.  The GC can have the maximum of statistics but the signal is difficult to disentangle from the background.
  On the opposite side the signal from dSphs can be a weak signal but with no background at all.
     \label{fig:fig1}
  }
\end{figure}

\section{Indirect Dark Matter searches}

Weakly Interacting Massive Particles (WIMPs)  can self-annihilate to produce prompt or secondary gamma rays during the annihilation. If WIMPs are produced thermally in the early Universe, then  the current velocity-averaged self-annihilation cross section has a natural value of  
${\langle \sigma_{ann} v \rangle}\sim$3$\cdot$10$^{-26}$cm${^3}$s$^{-1}$.
WIMP models, such as the supersymmetric neutralino, give predictions for gamma ray energy spectra from the annihilations, which are crucial inputs, together with the DM distribution in the observed target, to estimate prospects for the sensitivity of indirect searches.
 The expected DM annihilation gamma ray flux from a DM-dominated region depends on the particle physics and astrophysical (or $J$) factors: 

\begin{equation}
\Phi_{s}(\Delta\Omega)=\frac{1}{4\pi}\frac{\langle \sigma v \rangle}{2m^{2}_{DM}}\int^{E_{max}}_{E_{min}}\frac{dN_{\gamma}}{dE_{\gamma}}dE_{\gamma} \times J(\Delta\Omega),
\end{equation}
where $\langle \sigma v \rangle$ is the velocity-averaged self-annihilation cross section, $m_{DM}$ is the dark matter particle mass, $E_{min}$ and $E_{max}$ are the energy limits for the measurement and $\frac{dN_{\gamma}}{dE_{\gamma}}$ is the energy spectrum of the gamma rays produced in the annihilation (as, e.g., from \cite{Cirelli}). The products of DM annihilation are thought to come from decay and/or hadronization of the primary Standard Model (SM) particles: quark-antiquark, lepton and boson, and each channel is expected to have its own branching ratio.  The $J$ factor is the integral along the line of sight of the squared DM density profile of the given target integrated within an aperture angle, $\int_{\Delta\Omega}\text{d}\Omega\int_{l.o.s.}\rho^{2}_{DM}({r})\text{d}l$. Until recently, simulations used only cold dark matter (CDM), included only the gravitational force, and usually predicted the dark matter density to go approximately as 1/r towards the center of the dark matter halos. Standard parameterizations of these simulated dark matter halos are the Navarro, Frenk and White (NFW)\cite{NFW} and the Einasto\cite{Einasto} profiles. The latter one is moderately shallower on small spatial scales compared to the NFW profile.
N-body simulations showed dark matter profiles that can be both steeper and shallower. Steeper profiles are usually referred to as cuspy profiles. All the dark matter simulations agree on the main halo structure at large distances but the predictive power is limited by the spatial resolution of the simulation, and the shape and density of the profile in the inner part of the halo relies on extrapolation of the simulation prediction.

\begin{figure}
  \centering
  \includegraphics[width=0.49\textwidth]{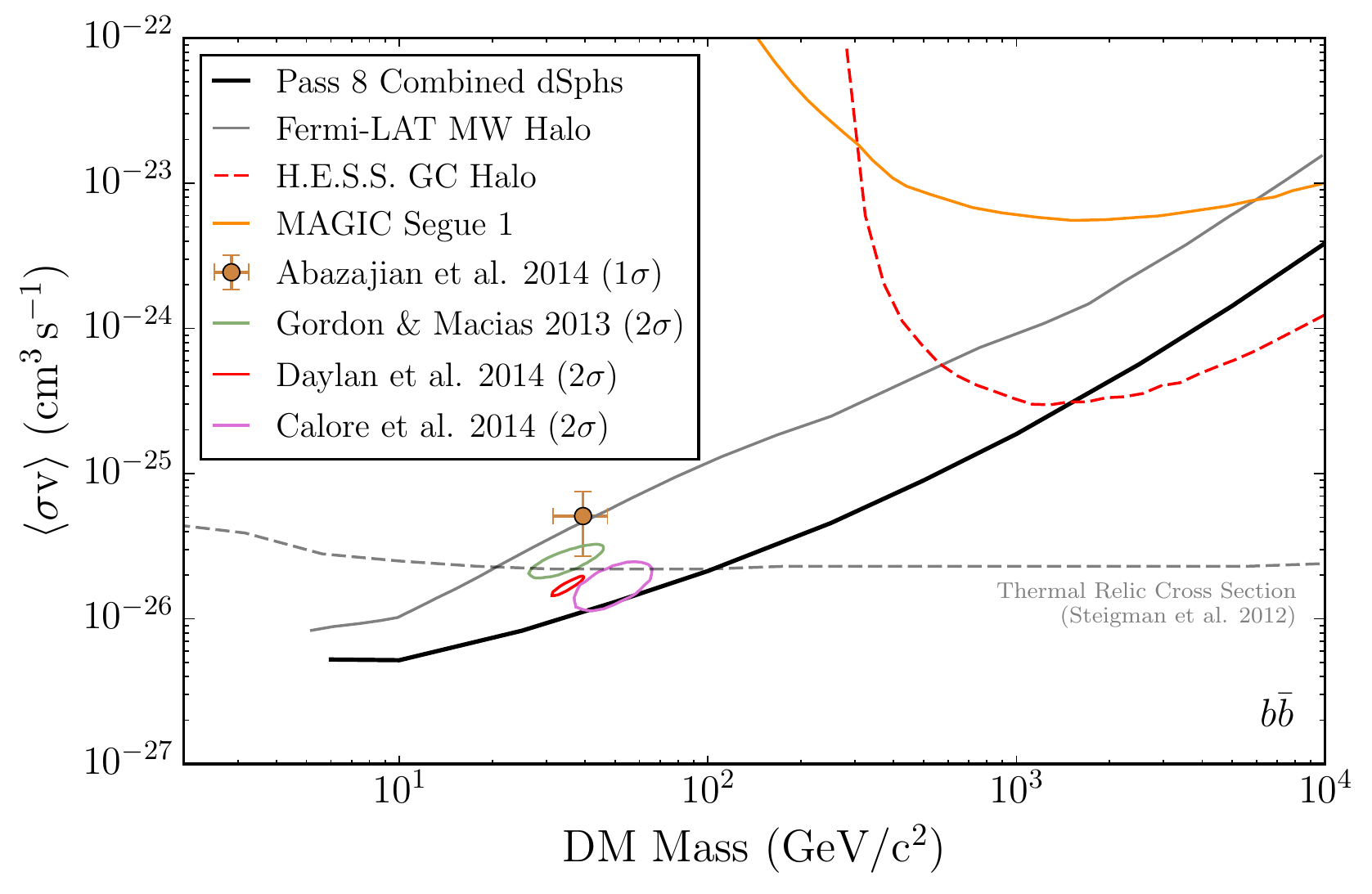}
  \includegraphics[width=0.49\textwidth]{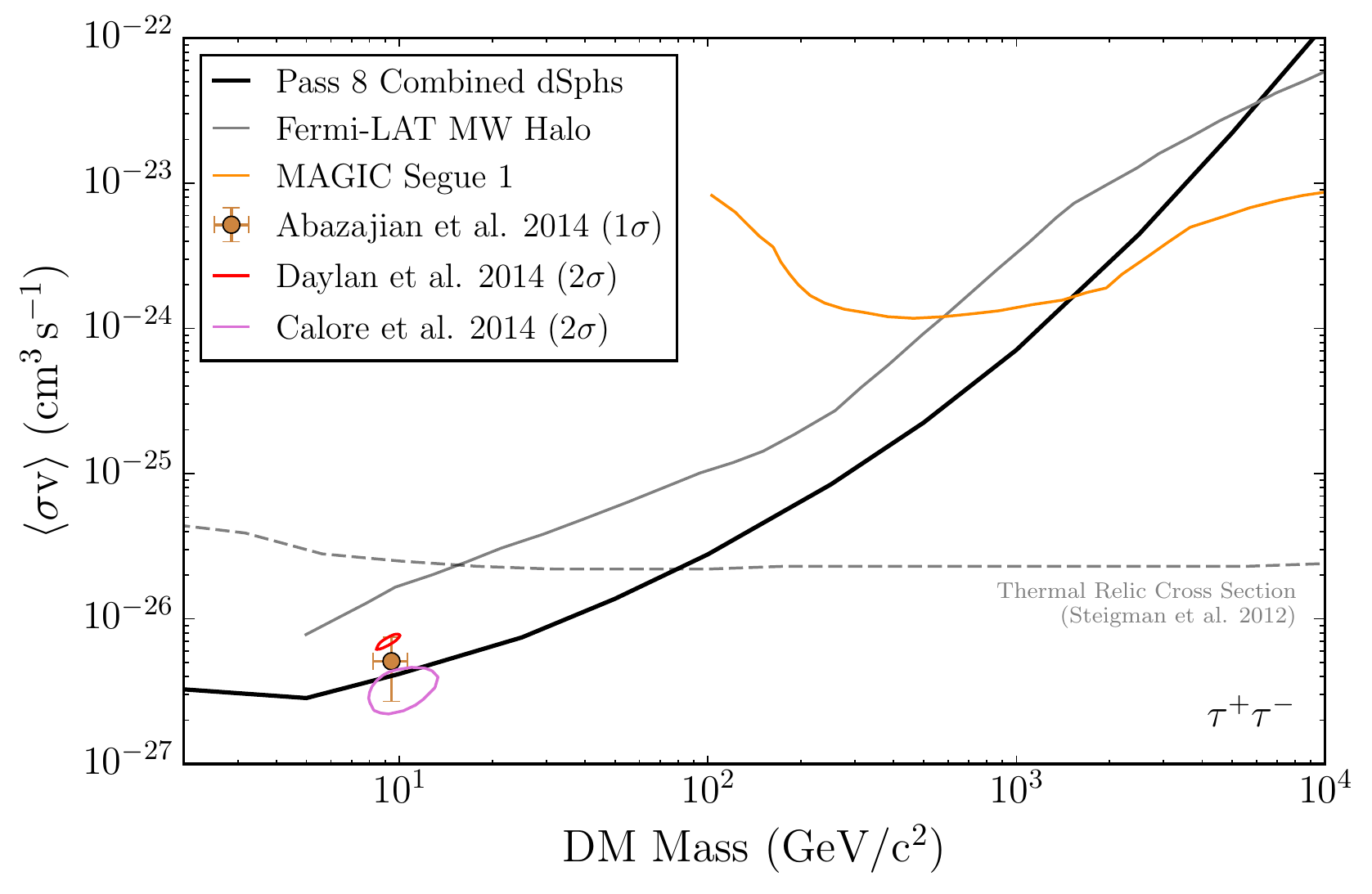}
  \caption{Comparison of constraints on the DM annihilation cross
    section for the $\overline b b$ (left) and $~\overline \tau \tau$ (right) channels \cite{Fermi_Dwarf} with previously published constraints from LAT analysis
    of the Milky Way halo ($3\sigma$ limit) \cite{fermidiffuse},
    112 hours of observations of the Galactic Center with H.E.S.S.
    \cite{Hess}, and 157.9 hours of observations of
    {Segue~1} with MAGIC \cite{Magic}.  Closed contours and
    the marker with error bars show the best-fit cross section and mass
    from several interpretations of the Galactic center excess
    \cite{Calore}.}\label{fig:fig2}
\end{figure}

\subsection{ Dwarf galaxies}
The dwarf spheroidal galaxies (dSphs) of the Milky Way are among the cleanest targets for indirect dark matter searches in gamma rays. They are systems with a very large mass/luminosity ratio (i.e., systems which are largely DM dominated). 
In a plot of signal strength versus robustness the dSphs are on the opposite side in respect to the GC ( see figure \ref{fig:fig1} ).
The Fermi-LAT detected no significant emission from any of such systems and the upper limits on the gamma ray flux allowed us to put very stringent constraints on the parameter space of well motivated WIMP models \cite{Dwarf}.
A combined likelihood analysis of the 10 most promising dwarf galaxies, based on 24 months of data and pushing the limits below the thermal WIMP cross section for low DM masses (below a few tens of GeV), has been  performed  \cite{Dwarf2}.
The derived 95\% C.L. upper limits on WIMP annihilation cross sections for different channels are shown in figure \ref{fig:fig2}. The  most generic cross section
 ($\sim$3 $\cdot$10$^{-26}$cm${^3}$s$^{-1}$ for a purely s-wave cross section) is plotted as a reference.  These results are obtained for NFW profiles \cite{NFW}, but for  cored dark matter profile the J-factors for most of the dSphs would either increase or not change much, so these results include J-factor uncertainties   \cite{Dwarf2}. 

\begin{figure}
  \centering
 \includegraphics[width=1.0\textwidth]{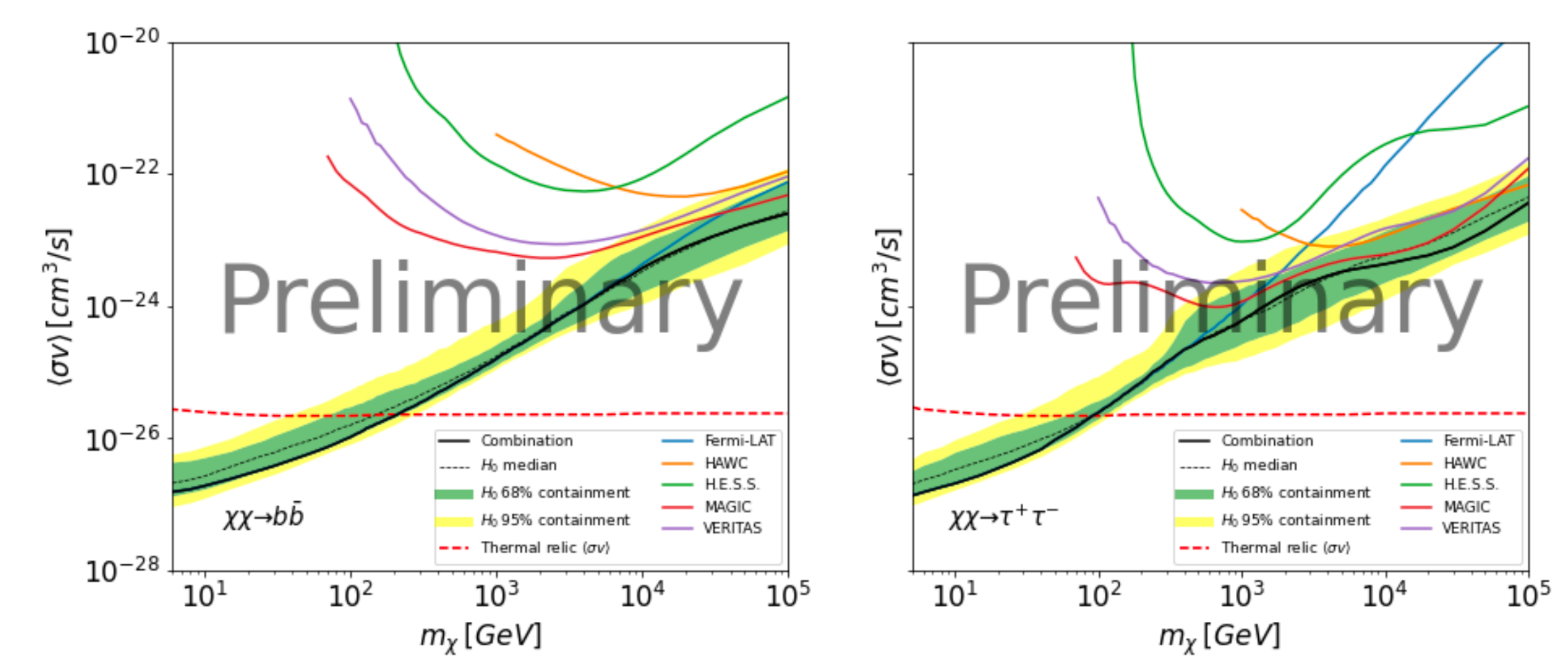}
  \caption{ Constraints on the DM annihilation cross
    section for the $\overline b b$ (left) and $~\overline \tau \tau$ (right) channels. The black solid line represents the observed combined limit, the black dashed line is the median of the null hypothesis corresponding to the expected limit, while the green and yellow bands show the 68 \% and 95 \% containment bands. Combined upper limits for each individual detector are also indicated as solid, colored lines \cite{Combined}.
    \label{fig:fig3}
  }
\end{figure}

The limits obtained with Fermi-LAT can be combined with the results of HAWC, H.E.S.S., MAGIC, and VERITAS
with the use of a joint maximum likelihood approach combining each experimentÕs individual analysis to derive more constraining upper limits on the WIMP DM self-annihilation cross section.
The results are shown in figure \ref{fig:fig3}.  Below $\sim$ 10 TeV the DM limits are largely dominated by Fermi-LAT  for the hadronic DM  $\overline b b$ channel,   then above ~10 TeV, the IACTs (H.E.S.S., MAGIC, and VERITAS) and HAWC take over. In the case of the leptonic channel, both the IACTs and HAWC contribute significantly to the DM limit from ~1 TeV to ~100 TeV.
With the present data   we are able to rule out
large parts of the parameter space  where the thermal relic density is
below the observed cosmological dark matter density and WIMPs  are
dominantly produced non-thermally, e.g. in models where supersymmetry
breaking occurs via anomaly mediation up to $\sim$ 100 GeV.

\subsection{ Galactic center}
The Galactic center (GC) is expected to be the strongest source of gamma rays from DM 
annihilation, due to its coincidence with the cusped part of the DM halo density profile \cite{dark1},  \cite{dark}, 
 but the region  is one of the richest in the gamma ray sky. Gamma ray emission in this direction includes the products of interactions between cosmic rays (CRs) with interstellar gas (from nucleon-nucleon inelastic collisions and electron/positron bremsstrahlung) and radiation fields (from inverse Compton scattering of electrons and positrons), as well as many individual sources such as pulsars, binary systems, and supernova remnants (SNRs).
A preliminary analysis of  Fermi-LAT observations of the GC region was presented in  \cite{F_sym},  \cite{Good}, 
\cite{GC_cim}  and then analysed in \cite{GC_art}.
These results produced a lot of activity outside the Fermi-LAT collaboration with claims of
evidence for dark matter in the Galactic Center (i.e.  \cite{Calore} \cite{Daylan} and references therein).
This possibility was already considered in the analysis of the EGRET galactic center excess \cite{dark}
with results similar to the analysis of the Fermi-LAT data  
but there are other possible explanations, e.g., past activity of the Galactic Center \cite{Petrovic},\cite{Carlson} or a population of millisecond pulsars around the Galactic Center  \cite{Lee}.
The Fermi-LAT Collaboration studied again the Galactic Center data in \cite{GC_Coll}  with the use of 6.5 yr of data 
with a characterisation of the uncertainty of the GC excess spectrum and morphology due to uncertainties in cosmic-ray source distributions and propagation, uncertainties in the distribution of interstellar gas in the Milky Way, and uncertainties due to a potential contribution from the Fermi bubbles with the conclusion that the nature of the GeV excess is still unclear and more studies are needed.  A combination of millisecond pulsars and DM 
annihilation is also possible and it is claimed to give a better fit at high energy \cite{Cholis}.
A new experiment with better angular resolution at low energies can help to disentangle the potential contribution from other astrophysical sources (for instance unresolved pulsars) and can help to find the cause of the effect \cite{Gamma-Light}, \cite{astrogam}.

\begin{figure}
  \centering
 \includegraphics[width=1.0\textwidth]{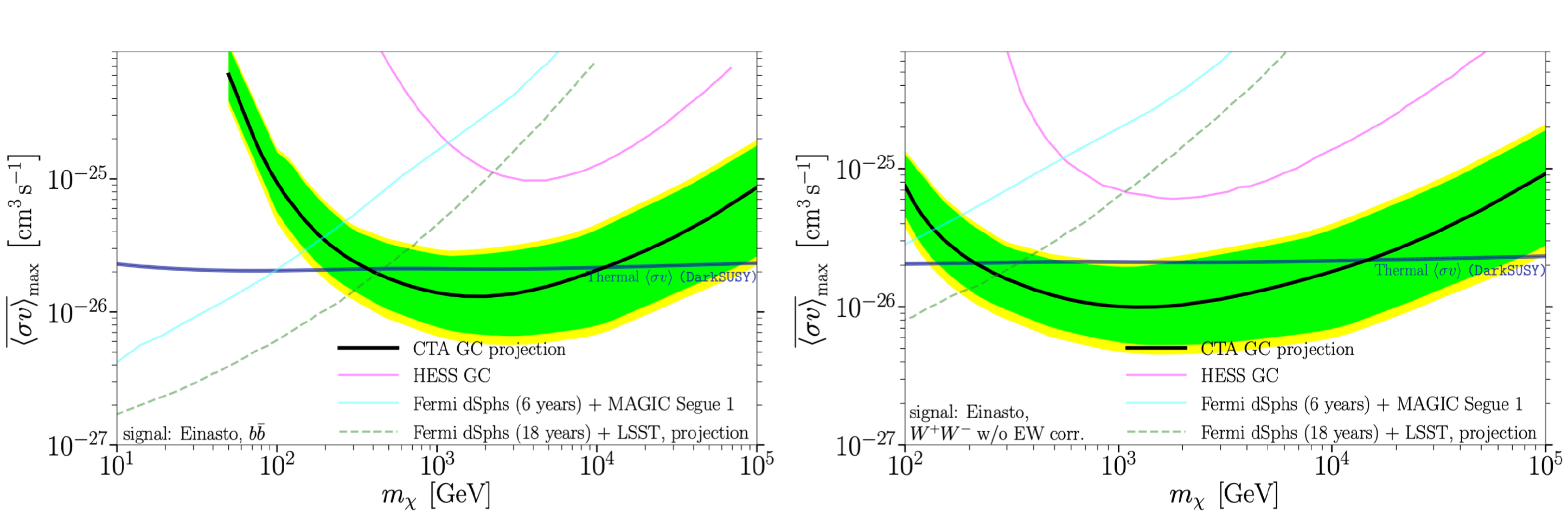}
  \caption{ The CTA sensitivity curves to a dark matter signal from the Galactic centre (black line) for $\overline b b $ (left) and $~W^+ W^-$ (right) channels foreseen for the Galactic centre Survey for a observation total time of  525 h, shown together with the current limits from Fermi-LAT observation of dSph galaxies (cyan)   \cite{Dwarf} and H.E.S.S. observations of the GC (purple)   \cite{Hess2} .  The dashed green is the projection  \cite{LSST} of the Fermi-LAT sensitivity where future dSphs discoveries with LSST are taken into account
  in the hypothesis of 18 years of observations.     \label{fig:fig4}
  }
\end{figure}

In the future Fermi-LAT should be able to observe the same sources for longer time and to 
observe new ones, so in principle it can make a discovery or put stronger limits.
This will complement the limits at higher energies that can be obtained with the CTA observation at the Galactic Center  (see figure \ref{fig:fig4} \cite{CTA_GC} ).
By observing the region around the Galactic Centre and by adopting dedicated observational strategies 
CTA will indeed reach the canonical velocity-averaged annihilation cross section for a dark matter mass in the range $\sim$ 200 GeV to 20 TeV,   something which is not possible with current instruments for any exposure time.
The observational strategy proposed for the CTA Dark Matter Programme is focused first on collecting a significant amount of data on the Galactic Centre. Complementary observations of a dSph galaxy will be conducted to extend the search and an article is in preparation  to define the best dSph to observe both in the northern and the southern skies \cite{CTA_dsph}. The Galactic Centre, Large Magellanic Cloud (LMC), and galaxy clusters are valuable targets both for dark matter searches and for studies of non-thermal processes in astrophysical sources. Data will be searched for continuum emission and line features, and strategies will be adopted according to the findings. Discoveries will modify any strategies defined a priori.   LMC, a nearby satellite galaxy at high Galactic latitude with the shape of a disk seen nearly face-on at a distance of only $\sim50$~kpc  with a large dark matter mass of $\sim 10^{10}$~M$_\odot$,  is an extended source for CTA.  Most of the emission lies within the CTA field of view, enabling the full galaxy to be observed in detail.  Like the Milky Way Galactic halo, astrophysical backgrounds are a significant challenge but the observation of LMC has the advantage that it is interesting not only for dark matter search but because it hosts many interesting astrophysical sources
 \cite{CTA_Science}. 

The observations of the best dwarf spheroidal galaxy will be started in the first year of the Dark Matter Programme. Any hints of dark matter signals or unknown sources would guide the plans for future observations. In the absence of signals, a programme of observation on the most promising dSph would be taken.
In conclusion, the WIMP paradigm, either through detection or non-detection will be significantly impacted upon during the first years of operation of CTA.

\section{Acknowledgments}
This work was conducted in the context of the CTA and Fermi
Dark Matter and New Physics Working Groups.
We gratefully acknowledge financial support from the agencies and organizations listed here: http://www.cta-observatory.org/consortium\_acknowledgments.

\end{document}